\documentclass[prd, preprint, 11pt]{revtex4-1}

\usepackage{amsmath}
\usepackage{amssymb}
\usepackage{setspace}
\usepackage{graphicx}
\usepackage{natbib}

\begin{document}


\begin{center}
 { \large {\bf Is quantum linear superposition an exact principle of nature?}}



\vskip 0.2 in

{\large{\bf Angelo Bassi$^a$, Tejinder  Singh$^b$ and Hendrik Ulbricht$^c$}}

\medskip
{\it $^a$Department of Physics, University of Trieste, Strada Costiera 11, 34151 Trieste, Italy}\\
{\it $^b$Tata Institute of Fundamental Research,}
{\it Homi Bhabha Road, Mumbai 400 005, India}\\
{\it $^c$School of Physics and Astronomy, University of Southampton, SO17 1BJ, UK}
\medskip

\end{center}

\centerline{\bf ABSTRACT}
\smallskip
\setstretch{1.1}
\noindent The principle of linear superposition is a hallmark of quantum theory. It has been confirmed experimentally for photons, electrons, neutrons, atoms, for molecules having masses up to ten thousand amu, and also in collective states such as SQUIDs and Bose-Einstein condensates.   However, the principle does not seem to hold for positions of large objects! Why for instance, a table is never found to be in two places at the same time?  One possible explanation for the absence of macroscopic superpositions is that quantum theory is an approximation to a stochastic nonlinear theory. This hypothesis may have its fundamental origins in gravitational physics, and is being put to test by modern ongoing experiments on matter wave interferometry.

\medskip

\centerline{{\bf {This essay received the Fourth Prize in the FQXi Essay Contest, 2012}}}




\setstretch{1.1}

\section{The absence of macroscopic superpositions}

\noindent In the year 1927, American physicists Clinton Davisson and Lester Germer performed an experiment at Bell Labs, in which they scattered a beam of electrons off the surface of a nickel plate. In doing so, they accidentally discovered that the scattered electrons exhibited a diffraction pattern analogous to what is seen in the Bragg diffraction of X-rays from a crystal. This experiment established the wave nature of electrons, and confirmed de Broglie's hypothesis of wave-particle duality. An electron can be in more than one position at the same time, and these different position states obey the principle of quantum linear superposition: the actual state of the electron is a linear sum of all the different position states.

The principle of linear superposition is the central tenet of quantum theory, an extremely successful theory for all observed microscopic phenomena. Along with the uncertainty principle, it provides the basis for a mathematical formulation of quantum theory, in which the dynamical evolution of a quantum system is described by the Schr\"{o}dinger equation for the wave function of the system. 

The experimental verification of linear superposition for electrons heralded a quest for a direct test of this principle for larger, composite particles and objects. Conceptually, the idea of the necessary experimental set up is encapsulated in the celebrated double slit interference experiment. A beam of identical particles is fired at a screen having two tiny slits separated by a distance of the order of the de Broglie wavelength of the particles, and an interference pattern is observed on a photographic plate at the far end of the screen. Such an experiment was successfully carried out for Helium ions, neutrons, atoms, and small molecules, establishing their wave nature and the validity of linear superposition.

When one considers doing such an interference experiment for bigger objects such as a beam of large molecules the technological challenges become enormous. The opening of a slit should be larger than the physical size of the molecule (say a few hundred nanometres) and the separation between slits should be smaller than the coherence length (few micrometres) ~\cite{Hornberger2011review}. The first experiment in this class was performed in 1999 in Vienna with $C_{60}$ molecules [fullerene, having nearly 700 nucleons per molecule] and the observed interference, as predicted by quantum theory, dramatically vindicated the validity of linear superposition for such large systems ~\cite{arndt1999wave}. Today, a little over a decade later, experimentalists have succeeded in pushing this limit to molecules with about 10,000 nucleons, and are aiming to push it higher to particles with a million nucleons in the next few years.  This is an extraordinary feat, considering the great difficulties involved in manouevering such large particles as they travel from the source, through the slits, and on to the detector where the interference pattern forms.

But will the principle of linear superposition continue to hold for larger and larger objects? The answer is indeed yes, according to the modern outlook towards quantum theory. The theory does not say that superposition should hold only in the microscopic world - in fact, a molecule with ten thousand nucleons, for which the principle has been confirmed,  isn't exactly microscopic!

However when we look at the day to day world around us linear superposition does not seem to hold! A table for instance is never found to be `here' and `there' at the same time. In other words, superposition of position states does not seem to hold for macroscopic objects. In fact already at the level of a dust grain, which we can easily see with the bare eye, and which has some $10^{18}$ nucleons, the principle breaks down. What could be going on in the experimentally untested desert bewteen $10^{4}$ nucleons, where linear superposition is valid, and $10^{18}$ nucleons, where it is not valid?

The observed absence of macroscopic superpositions is the source of the quantum measurement problem, as illustrated by the double slit experiment for a beam of electrons. The interference pattern which forms when an electron passes through both the slits is destroyed when a detector [measuring apparatus] is placed close to and behind one of the slits. The detector clicks (does not click) with a probability proportional to the square of the complex amplitude for the electron to pass through the upper slit (lower slit). This is the so-called Born probability rule. The presence of the detector breaks the superposition of the two states `electron going through upper slit' and `electron going through lower slit' and the interference pattern is lost. One says that the quantum state of the electron has collapsed, either to `electron going through upper slit only', or to `electron going through lower slit only'. The collapse of the superposition of the two electron states [up and down], and the emergence of probabilities contradict the Schr\"{o}dinger equation, since this equation is deterministic and linear. It predicts that a superposition should be preserved during the measurement process, that there should be no probabilities, and that position superpositions should be observed for macroscopic objects as well, such as a measuring apparatus.

Why are superpositions of different position states for macroscopic objects not observed, in contrast with what quantum theory predicts? Possible explanations include reinterpretations of quantum theory such as the many worlds interpretation or consistent histories, and mathematical reformulations of the theory such as Bohmian mechanics. By introducing one or more additional assumptions, these reinterpretations and reformulations ensure that macroscopic superpositions will not be observed, while making the same experimental predictions as quantum theory.  

\section{Quantum theory is approximate}

Here we explore another possibility, which has been investigated extensively during the last three decades or so ~\cite{Adler:09, Bassi:03, RMP:2012}. What if quantum linear superposition is an approximate principle of nature? This means the following: consider an object (say an atom, a molecule, or a table) which consists of $N$ nucleons and is in a superposition of two quantum states of an observable one is trying to measure (for example spin or position). According to quantum theory the life time of such a superposition is infinite! However, no experiment to date rules out the possibility that the lifetime $\tau$ of such a superposition is  finite, and is a monotonically decreasing function of $N$.  The dependence of $\tau$ on $N$ should be such that for microscopic systems such as atoms the superposition life time is astronomically large [longer than the age of the Universe]. Furthermore, for large objects such as a table or a macroscopic apparatus the superposition life time should be so small that linear superposition is simply not observed on the scale of current experiments.

Somewhere in between the microworld and the macroworld, i.e. between objects of $10^{4}$ nucleons and $10^{18}$ nucleons, the superposition lifetime would be neither too large, nor too small, but just right enough to be detectable in the laboratory. In principle, this could be achieved as follows. Suppose one has prepared in a controlled manner a beam of very large identical molecules which are such that the superposition life time between two different position states is $\tau$. Let such a beam be passed through two slits in a double slit experiment, thereby creating a superposition of two different position states.  Let the distance of the photographic plate from the slits be large enough that the time of travel of a molecule from the slits to the plate far exceeds $\tau$. Then the superposition will exponentially decay before the molecule reaches the plate, and no interference pattern will be seen. And this will happen even though no detector has been placed behind either of the slits!

The requirements on a mathematically consistent generalization of quantum theory in which superposition is an approximate principle are extremely stringent. To begin with the theory should be nonlinear: superposition of two allowed quantum states of the system should not be a stable allowed state. And yet for microsystems this theory should be in excellent agreement with standard quantum theory so that results of all known experiments are reproduced. The nonlinear process responsible for the breakdown of superposition should be stochastic (i.e. random) in nature because the outcome of a particular quantum measurement cannot be predicted. All the same, the nonlinear process should be such that a collection of outcomes of repeated measurements obey the Born probability rule. Another reason the nonlinear effect should be stochastic is that this helps avoid faster-than-light signalling. A deterministic (i.e. non-stochastic) nonlinear quantum mechanics is known to allow for superluminal communication. The nonlinear mechanism should be amplifying in nature, being negligible for microscopic systems, but becoming more and more important for large many particle systems, so that breakdown of superposition becomes more and more effective. 

It is a remarkable achievement that such a theory could be developed, and was given shape in the eighties by physicists Ghirardi, Pearle, Rimini and Weber ~\cite{Ghirardi:86, Ghirardi2:90}. This theory, which has come to be known as Continuous Spontaneous Localization [CSL] has been extensively investigated in the following two decades, with regard to its properties and solutions, constraints from laboratory and astrophysical observations, and direct tests of its predictions in the laboratory ~\cite{RMP:2012}. It is perhaps the only well studied generalization of quantum theory whose experimental predictions differ markedly from those of quantum theory in the macroworld  and which hence provides an excellent benchmark against which the accuracy of quantum theory can be tested.

In its original form [known as the GRW model], the CSL model is based on the following two principles:

1. Given a system of $n$ distinguishable particles, each particle experiences a sudden spontaneous localization (i.e. collapse of position state) with a mean rate $\lambda$,  to a spatial region of extent $r_C$.  

2. In the time interval between two successive spontaneous localizations, the system evolves according to the  Schr\"{o}dinger equation.

 Thus two new fundamental constants of nature have been introduced, and these are assumed to take definite numerical values, in order to successfully reproduce the observed features of the microscopic and the macroscopic world. The constant $\lambda$, assumed to be of the order $10^{-16}$ sec$^{-1}$, determines the rate of spontaneous localization for a single particle. Amplification is achieved by showing that for a composite object of $n$ particles, the collapse rate is $(\lambda n)^{-1}$ seconds. The second constant $r_C$ is a length scale assumed to be about $10^{-5}$ cm, and indicates that a widely spread quantum state collapses to a size of about $r_C$ during localization.

In its modern version, the CSL model consists of coupling a randomly fluctuating classical field with the particle number density operator of a quantum system, so as to produce collapse towards spatially localized
states. The collapse process is continuous and dynamics is described by a single nonlinear stochastic differential equation which contains both aspects: standard Schr\"{o}dinger evolution as well the nonlinear stochastic terms which result in collapse and breakdown of superposition. The fundamental constants $\lambda$ and $r_C$ continue to be retained. Today, experiments are being devised to test CSL and are approaching the ballpark range where the preferred values of $\lambda$ and $r_C$ will be confirmed or ruled out. 

The CSL dynamical equation can be used to show that macroscopic superpositions last for extremely short time intervals, thus explaining why such position superpositions are not observed. Similarly the superposition lifetime for microstates is extremely large. The same CSL equation elegantly solves the measurement problem. Suppose a microscopic quantum system is prepared in a superposition of two quantum states, say A and B,  of an observable which is to be measured. The CSL equation for this system is essentially the same as the Schr\"{o}dinger equation, the nonlinear effect being negligible, and evolution proceeds in the standard deterministic linear fashion, preserving superposition of the states A and B.  When the quantum system comes into interaction with the measuring apparatus, which let us say is characterized by a pointer, one must now write down the CSL equation for the combined system consisting of the quantum system and the measuring apparatus, and now the nonlinear stochastic terms come into play. The state A would cause the pointer to be at a position $P_1$ and the state B would cause the pointer to be at a different position $P_2$. The superposition of A and B would cause the pointer to be in superposition of the macroscopically different states $P_1$ and $P_2$. This however is not permitted by CSL, since the pointer is macroscopic: the superposition very quickly decays to the pointer state $P_1$ or $P_2$. This is precisely what is interpreted as the collapse of the superposed quantum state to either A or B. Furthermore, the CSL equation shows that the outcome of an individual measurement is random, but repeated measurements are shown to result in outcomes $P_1$ or $P_2$ with relative frequency as given by the Born probability rule ~\cite{Bassi:07}.

\section{Testing the idea with experiments}

The nonlinear stochastic modification of the Schr\"{o}dinger equation and the introduction of the constants $\lambda$ and $r_C$ modify the predictions of quantum theory for various standard results from known experiments and astrophysics. This allows bounds to be put on these constants. These include bounds coming from  decay of supercurrents in SQUIDS, spontaneous X-ray emission from Germanium, absence of proton decay, heating of the intergalactic medium, dissociation of cosmic hydrogen, heating of interstellar dust grains, and temperature distortions in the cosmic microwave background.  A novelty of the CSL mechanism is that the process of spontaneous localization produces a very tiny increase in the energy of the localized particle, thus violating energy conservation. Some of the bounds on the fundamental constants come from the lack of observation of any such violation.  Using other arguments based on latent image formation in photography ~\cite{Adler3:07} and formation of image in the human eye ~\cite{Bassi2:10}, it has been suggested that the theoretical value of $\lambda$ should be as high as $10^{-8}$ sec$^{-1}$. We thus already see a fascinating debate taking place between theory and experiment: the principle of quantum linear superposition is being confronted by experiment in the true sense of the word.

However the most direct tests of CSL and  linear superposition will come from interference experiments for large objects. Unlike quantum theory, CSL predicts that under certain suitable conditions an interference pattern will not be seen. Thus an accurate experiment carried out under these conditions will definitely establish whether quantum linear superposition is an exact or approximate principle. If interference is seen with a large molecule, it sets an upper bound on the value of $\lambda$. This experimental field, known as matter wave interferometry, has made great strides in recent years, and is one of the most important sources for testing proposed alternatives to quantum theory, such as CSL ~\cite{Feldmann:11, Hornberger2011review, nimmrichter2011testing, Romero2012decoherence, Arndt2011focusissue}.

The experiments involve overcoming technical challenges to prepare intense beams of molecules in gas phase, to preserve spatial and temporal coherence of the beam, and to detect the particles efficiently. Beam splitters for molecules are typically highly ordered periodic diffraction gratings of  nanowires made from metal or semiconductors, or they are standing light fields realized using the so-called Kapitza-Dirac effect.  The 1999 Vienna experiment with $C_{60}$ fullerene used far-field Fraunhofer molecular diffraction with a grating constant of 100 nanometres. For larger molecules, where it becomes imperative to effectively increase the beam intensity, more promising results have been achieved through near field interference using the Talbot-Lau interferometer [TLI].

\begin{figure}[!htb]
  \centerline{\includegraphics[totalheight=0.45\textheight, width=.7\textwidth]{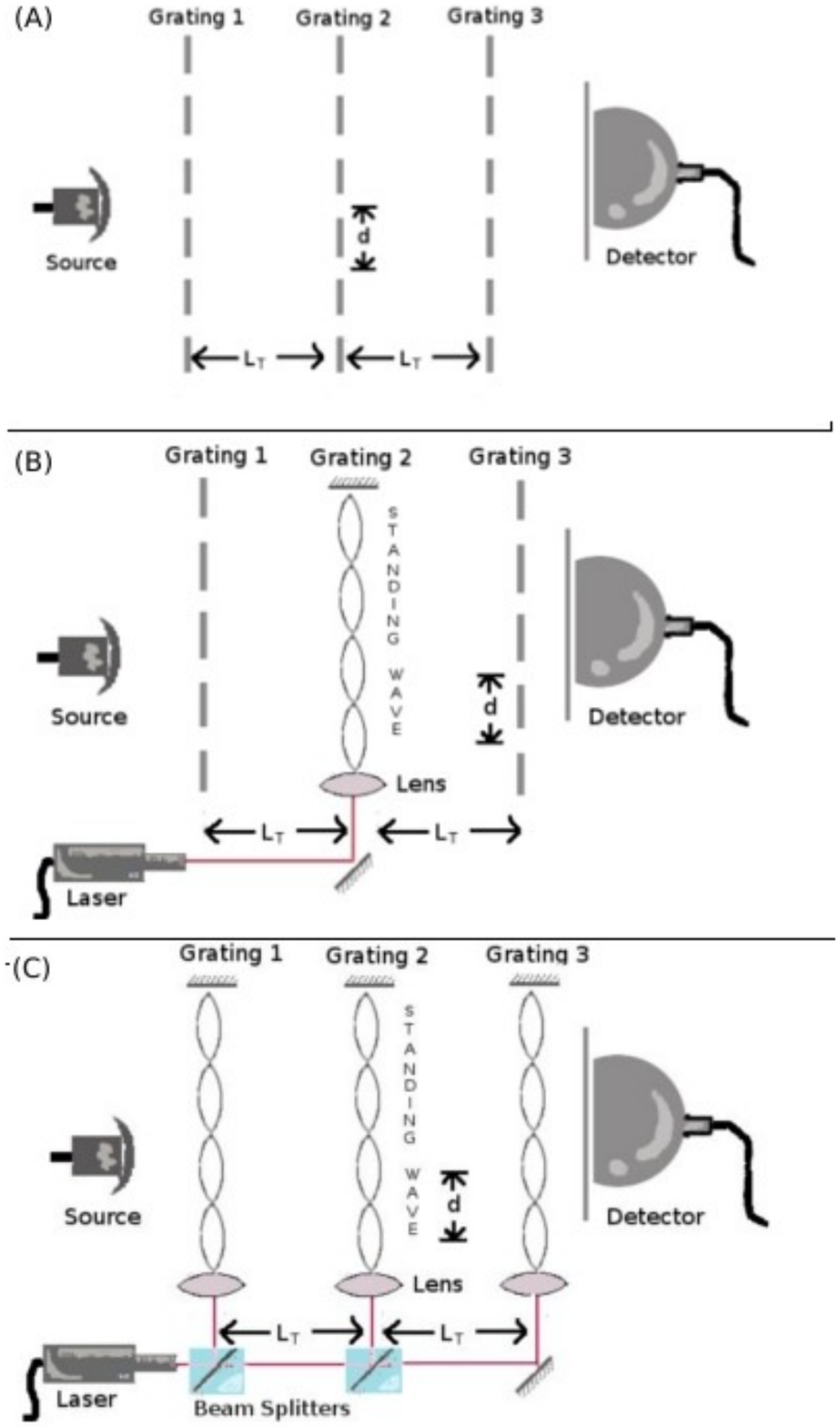}}%
 \caption{Different configurations of the Talbot-Lau interferometer are shown. (A) Three material grating (B) Kapitza-Dirac-Talbot-Lau interferometer (C) Optical Time-Domain Ionizing Matter Interferometer
[OTIMA]. Figure Courtesy: Kinjalk Lochan. Figure Source ~\cite{RMP:2012}. }\label{fig.TLI}
\end{figure}

A TLI operating in the near-field (i.e. the spatial period of the gratings and the interference pattern are on the same scale) was specifically invented to deal with beams of low intensity and low collimation in interference experiments ~\cite{clauser1992new}.  A three grating TLI operates with weakly collimated beams: the first grating prepares beam coherence and imprints a spatial structure on the beam. The second grating is thus simultaneously illuminated by some $10^{4}$ individual coherent molecular beams and creates a self-image on a third grating, on which the interference pattern forms. Effectively, the number of molecules contributing to the final interference pattern is multiplied by the number of illuminating slits of the first grating, and all the coherent beams from the $10^{4}$ source slits are incoherently summed to contribute to the same interference pattern. More recently, a modified version, known as the Kapitza-Dirac TLI ~\cite{hornberger2009theory} has been employed, in which the second - the diffraction grating is replaced by an optical phase grating [Fig 1]. Molecules are diffracted at periodic optical potentials due to the 
Kapitza-Dirac effect. The KDTLI has been used to demonstrate interference with a molecule
known as perfluoro-alkylated $C_{60}$, having a mass 7,000 amu ~\cite{Gerlich2011}. This is the largest molecule on which a successful matter wave experiment has been carried out so far, and sets an upper bound of $10^{-5}$ sec$^{-1}$
on the CSL parameter $\lambda$. This is only three orders of magnitude away from the theoretical value $10^{-8}$ for $\lambda$ predicted by Adler, and this latter value could be confronted by an experiment with molecules having a mass of about 500,000 amu.

However, for molecules above $10^{5}$ amu, their electromagnetic interactions with the material gratings  disable the interference pattern, and new technologies must be sought to efficiently control and manipulate the center of mass motion of heavy particles. Experiments are performed in ultra-high vacuum conditions to prevent decoherence by collisions. Beams should be slow, so that the de Broglie wavelength is not too low, they should be highly collimated and should have a high phase space density. These features can be achieved through promising cooling techniques currently under development. Another important aspect is whether to use neutral or charged particles. All large particle interference experiments thus far have been performed with neutrals - they have the advantage that they suffer lesser decohering effects from interaction with the environment. On the other hand charged particles are easier to manipulate and control, especially while preparing coherent particle beams of heavy molecules 
~\cite{RMP:2012}.

A clever new proposal combines the best of both worlds: manipulation of charged particles during the preparation of the beam, and interference after neutralization. This novel three light grating TLI aims towards the interference of particles up to $10^{9}$ amu and is known as the optical time-domain matter-wave (OTIMA) interferometer ~\cite{Nimmrichter2011concept}.  Charged particles will be provided by a mass filtered cluster aggregration source. The charged clusters are neutralized at the first grating using light-matter effects, diffracted at the second grating, and ionized again for detection at the third grating. 

An alternative approach to testing quantum superposition and CSL is optomechanics, which involves coupling micromechanical devices to light fields with which they interact ~\cite{Bose1997, Marshall:03}. A tiny mechanical object such as a micromirror is cooled to extremely low temperatures and prepared to be, say, in a quantum superposition of the vibrational ground state and first excited state. This superposed state is  coupled to a sensitive optical interferometer: if the superposed state decays in a certain time, as predicted by CSL, the optical response of the interferometer will be different, as compared to when the superposition does not decay [Fig. 2]. Optomechanics aims to test superposition for very massive particles in the range $10^{6}$ to 
$10^{15}$ amu,  but the vibrational amplitude is very small compared to the desired amplitude, which is the fundamental length scale $r_C$ of CSL. This makes it a challenge for optomechanics to reach the expected regime where new physics is expected ~\cite{revKippenberg2008, revAspelmeyer2010, chan2011laser, o2010quantum}.

\begin{figure}[!htb]
  \centerline{\includegraphics[width=10.0cm]{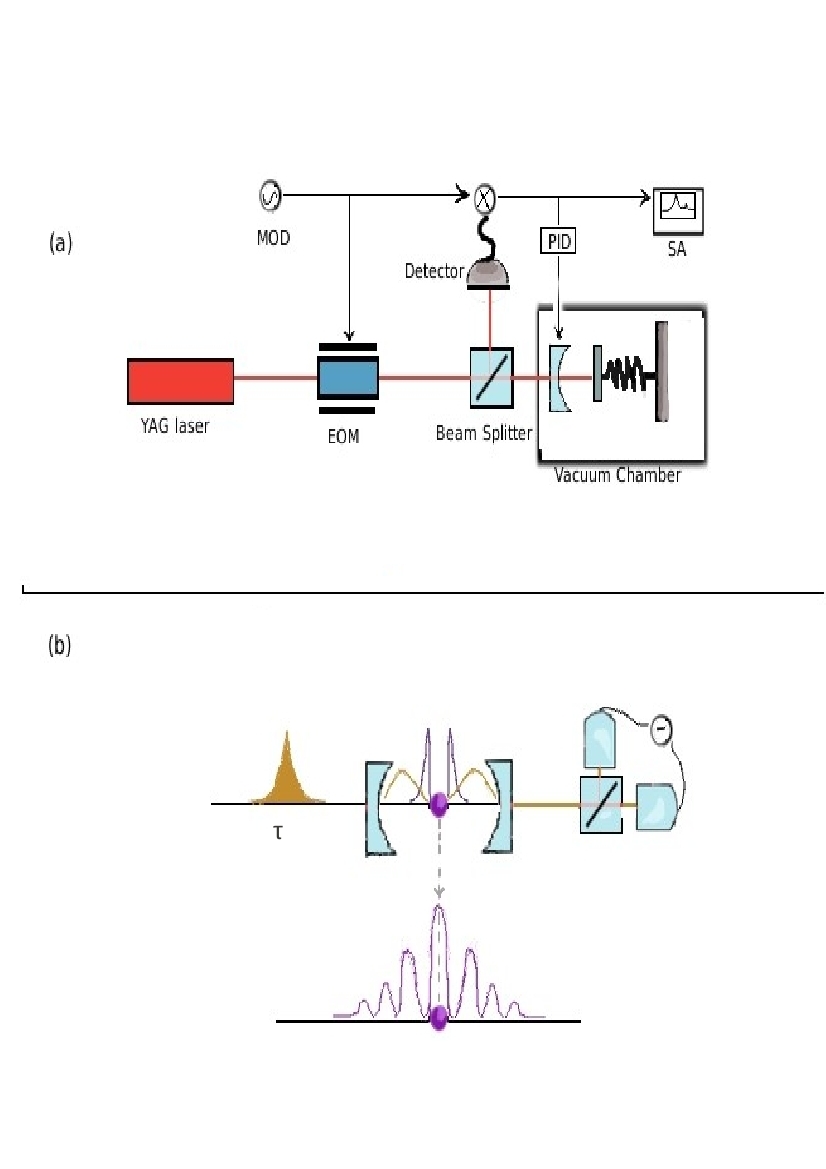}}
  \caption{Optomechanics: [a]  Prototype of optomechanically cooled cantilever. Quantum optical detection techniques enable the sensitive read out of vibrations as they couple to light fields.
  [b] Mechanical Resonator Interference in a Double Slit (MERID). The centre of mass motion of a single optically trapped nanoparticle is first cooled and then superimposed by an optical double potential. The interference pattern evolves in free fall after switching off the trapping field. Figure Courtesy: Kinjalk Lochan. Figure Source ~\cite{RMP:2012}.} \label{fig.optomech}
\end{figure}

However, very promising progress can be expected by combining techniques from optomechanics and matter wave interferometry. Massive particles trapped in optical traps are analogues of optomechanical systems, with the significant difference being that by suitable choice of the trapping potentials, superposition of position states can be created with a separation comparable to $r_C$. After the superposed state has been created the trapping potential is swiched off; the particles are allowed to fall freely, and their spatial density distribution can be studied to verify the existence of an interference pattern. Such an experiment can be carried out for very massive objects (such as polystyrene beads of about 30 nanometer diameter and mass $10^{6}$ amu). Known as MERID (mechanical resonator interference in a double slit), it is a promising future technique for testing the CSL model ~\cite{romero2011large}. It can be expected that within the next two decades matter wave interferometry and optomechanics experiments will reach the ballpark of $10^{6}$ amu to $10^{9}$ amu where predictions of CSL will differ from quantum theory sufficiently enough for them to be discriminated in the laboratory.

\section{Why is quantum theory approximate?}

Continuous Spontaneous Localization has been proposed as a phenomenological modification of quantum theory which is successful in explaining the observed absence of macroscopic superpositions, and the Born  rule. It is beyond doubt though that underlying CSL there ought to be deep physical principles which compel us to accept such a radical modification of quantum theory. Fascinating progress has been taking place towards unravelling these underlying principles, mainly along three different directions, all of which point to an involvement of gravity, and a revision of our understanding of spacetime structure at the deepest level.

It has been suggested independently by Karolyhazy et al. ~\cite{Karolyhazi:86}, Diosi ~\cite{Diosi:87} and Penrose ~\cite{Penrose:96} that gravity is responsible for the absence of macroscopic superpositions. While their arguments differ somewhat at the starting point they  all come to approximately the same conclusion that the absence of superpositions will become apparent around $10^{6} - 10^{9}$ amu, a range that agrees well with the prediction of CSL! The key idea is that every object
obeys the uncertainty principle and hence there is an intrinsic minimal fluctuation in the spacetime geometry produced by it. When the quantum state describing this object propagates in this fluctuating spacetime, it loses spatial coherence beyond a certain length scale after a critical time. Such a loss of spatial coherence is an indicator of breakdown of superposition of different position states. This length and time scale is shown to be astronomically large for microscopic particles, and very small for macroscopic objects, thus demonstrating macroscopic localization. There is tantalizing evidence in the literature ~\cite{Diosi:89} that the stochastic mechanism of CSL is provided by spacetime fluctuations, and that the fundamental constants $\lambda$ and $r_C$ derive from gravity. An ongoing optomechanics experiment is specifically devoted to testing the role of gravity in causing breakdown of superposition ~\cite{Marshall:03}.

A second remarkable line of development, known as Trace Dynamics, has come from the work of Stephen Adler and collaborators ~\cite{Adler:04} who suggest that it is unsatisfactory to arrive at a quantum theory by quantizing its very own limit, namely classical dynamics. Instead, quantum theory is derived as an equilibrium statistical thermodynamics of an underlying unitarily invariant classical theory of matrix dynamics. It is fascinating that the consideration of Brownian motion fluctuations around the equilibrium theory provides a nonlinear stochastic modification of the Schr\"{o}dinger equation of the kind proposed in CSL.  This is another clue that the absence of macroscopic superpositions may have to do with quantum theory being an approximation to a deeper theory.

Thirdly, the absence of macroscopic superpositions is possibly related to another deep rooted but little appreciated incompleteness of quantum theory. In order to describe dynamical evolution the theory depends on an external classical time, which is part of a classical spacetime geometry. However, such a geometry is produced by classical bodies, which are again a limiting case of quantum objects! This is yet another sense in which the theory depends on its classical limit. There hence ought to exist a reformulation of quantum theory which does not depend on a classical time ~\cite{Singh:2009}. Such a reformulation has been developed by borrowing ideas from Trace Dynamics and there is evidence that there exist stochastic fluctuations around such a reformulated theory, which have a CSL type structure, and are responsible for the emergence of a classical spacetime and breakdown of superposition in the macroscopic world ~\cite{Singh:2011, 
Lochan-Singh:2011, Lochan:2012}. 

The CSL model, as it is known today, is nonrelativistic in character, which means that it is a stochastic generalization of the nonrelativisitic Schr\"{o}dinger equation. It is intriguing that the model has thus far resisted attempts at a relativisitic generalization. On the other hand it is known that the 
collapse of the wave function during a quantum measurement is instantaneous and non-local. This has been confirmed by experimental verification of Bell's inequalities. Thus it is certainly true that there is a need for  reconciliation between CSL induced localization, and the causal structure of spacetime dictated by special relativity. In a remarkable recent paper ~\cite{Brukner:11} it has been shown that if one does not assume a predefined global causal order, there are multipartite quantum correlations which cannot be understood in terms of definite causal order and spacetime may emerge from a more fundamental structure in a quantum to classical transition.

For nearly a century the absence of macroscopic superpositions, in stark contradiction with what quantum theory predicts, has confounded physicists, and led Schr\"{o}dinger to formulate his famous cat paradox. The quantum measurement problem, a direct consequence of this contradiction, has been debated endlessly, and very many solutions proposed, by physicists as well as philosophers. However up until recent times the debate has remained largely theoretical, for no experiment has ever challenged the phenomenal successes of quantum theory. Times have changed now. There is a phenomenological model which proposes that quantum linear superposition is an approximate principle; there are serious underlying theoretical reasons which suggest why this should be so, and most importantly, experiments and technology have now reached a stage where this new idea is being directly tested in the laboratory. Perhaps after all it will be shown that the assumption that linear superposition is exact is a wrong assumption. We will then have nothing short of a revolution, which will have been thrust on us by experiments which disagree with quantum mechanics, thus forcing a radical rethink of how we comprehend quantum theory, and the structure of spacetime.

\medskip

\noindent {\bf Acknowledgement:} This work is supported by a grant from the John Templeton Foundation.


\newpage 

\centerline{\bf REFERENCES}

\bibliography{biblioqmts3}

\newpage

\setstretch{1.0}

\centerline{\underline{\bf TECHNICAL ENDNOTES} \qquad ({\it for details see ~\cite{RMP:2012}})}
\bigskip
\centerline{\underline{\bf The Physics of Continuous Spontaneous Localization}}
\medskip
\noindent The essential physics of the CSL model can be described by a simpler model, known as QMUPL (Quantum Mechanics with Universal Position Localization), whose dynamics is given by the following stochastic nonlinear Schr\"{o}dinger equation
\begin{equation} \label{eq:qmupl1}
d \psi_t  =  \left[ -\frac{i}{\hbar} H dt + \sqrt{\lambda} (q - \langle q \rangle_t) dW_t  
 - \frac{\lambda}{2} (q - \langle q \rangle_t)^2 dt \right] \psi_t,
\end{equation}
where $q$ is the position operator of the particle, $\langle q \rangle_t \equiv \langle \psi_t | q | \psi_t \rangle$ is the quantum expectation, and $W_t$ is a standard Wiener process which encodes the stochastic effect. Evidently, the stochastic term is nonlinear and also nonunitary. The collapse constant $\lambda$ sets the strength of the collapse mechanics, and it is chosen proportional to the mass $m$ of the particle according to the formula
$
\lambda = \frac{m}{m_0}\; \lambda_0,
$
where $m_0$ is the nucleon's mass and $\lambda_0$ measures the collapse strength. If we take $\lambda_0\simeq 10^{-2}$ m$^{-2}$ sec$^{-1}$ the strength of the collapse model corresponds to the CSL model in the appropriate limit.

The above dynamical equation can be used to prove position localization; consider for simplicity a free particle  $(H=p^2/2m)$ in the gaussian state (analysis can be generalized to other cases):
\begin{equation} \label{gsol}
\psi_{t}(x) = \makebox{exp}\left[ - a_{t} (x -
\overline{x}_{t})^2 + i \overline{k}_{t}x + \gamma_{t}\right].
\end{equation}
By substituting this in the stochastic equation it can be proved that the spreads in position and momentum
\begin{equation}
\sigma_{q}(t)  \equiv  \frac{1}{2}\sqrt{\frac{1}{a_{t}^{\makebox{\tiny
R}}}};\qquad
\sigma_{p}(t)  \equiv  \hbar\,\sqrt{\frac{(a_{t}^{\makebox{\tiny R}})^2 +
(a_{t}^{\makebox{\tiny I}})^2}{a_{t}^{\makebox{\tiny R}}}},
\end{equation}
do not increase indefinitely but reach asymptotic values given by 
\begin{equation} \label{aval1}
\sigma_{q}(\infty) = \sqrt{\frac{\hbar}{m\omega}} \simeq
\left( 10^{-15} \sqrt{\frac{\makebox{Kg}}{m}}\; \right)\,
\makebox{m}, \qquad
\sigma_{p}(\infty) = \sqrt{\frac{\hbar m\omega}{2}}
\simeq \left( 10^{-19} \sqrt{\frac{m}{\makebox{Kg}}}\,
\right)\, \frac{\makebox{Kg m}}{\makebox{sec}},
\end{equation}
such that:
$
\sigma_{q}(\infty)\, \sigma_{p}(\infty) \; = \;
{\hbar}/{\sqrt{2}}
$
which corresponds to almost the minimum allowed by Heisenberg's
uncertainty relations. Here, $\omega \; = \; 2\,\sqrt{{\hbar \lambda_{0}}/{m_{0}}} \; \simeq
\; 10^{-5} \; \makebox{s$^{-1}$}.$

Evidently, the spread in position does not increase indefinitely, but stabilizes to a finite value, which is a compromise between the Schr\"odinger's dynamics, which spreads the wave function out in space, and the collapse dynamics, which shrinks it in space. For microscopic systems, this value is still relatively large ($\sigma_{q}(\infty) \sim 1$m for an electron, and $\sim 1$mm for a $C_{60}$ molecule containing some 1000 nucleons), such as to guarantee that in all standard experiments---in particular, diffraction experiments---one observes interference effects. For macroscopic objects instead, the spread is very small ($\sigma_{q}(\infty) \sim 3 \times 10^{-14}$m, for a 1g object), so small that for all practical purposes the wave function behaves like a point-like system. This is how localization models are able to accommodate both the ``wavy'' nature of quantum systems and the ``particle'' nature of classical objects, within one single dynamical framework. 

The same stochastic differential equation solves the quantum measurement problem and explains the Born probability rule without any additional assumptions. For illustration, consider a two state microscopic quantum system ${\cal S}$ described by the initial state
\begin{equation}
c_{+}|+\rangle + c_{-} |-\rangle
\end{equation}
interacting with a measuring apparatus ${\cal A}$ described by the position of a pointer which is initially in a `ready' state $\phi_{0}$ and which measures some observable $O$, say spin, associated with the initial quantum state of ${\cal S}$. As we have seen above, the pointer being macroscopic [for definiteness assume its mass to be 1 gram], is localized in a gaussian state $\phi^{G}$, so that the initial composite state of the system and apparatus is given by
\begin{equation} \label{eq:10}
\Psi_{0} = \left[ c_{+} |+\rangle + c_{-} |+\rangle \right]
\otimes \phi^{\mathrm{G}}.
\end{equation}
According to standard quantum theory, the interaction leads to the following type of evolution:
\begin{equation} \label{eq:6}
{\left[ c_{+} |+\rangle + c_{-} |-\rangle \right] \otimes \phi^{G}} \qquad
 \mapsto  \qquad  c_{+} |+\rangle \otimes \phi_{+}  + c_{-} |-\rangle
\otimes \phi_{-},
\end{equation}
where $\phi_{+}$ and $\phi_{-}$  are the final  pointer states corresponding to the system being in the collapsed state $|+\rangle$ or $|-\rangle$ respectively. While quantum theory explains the transition from the entangled state (\ref{eq:6}) to one of the collapsed alternatives by invoking a new interpretation or reformulation, the same is achieved dynamically by the stochastic nonlinear theory given by (\ref{eq:qmupl1}).

 It can be proved from (\ref{eq:qmupl1}) that the initial state (\ref{eq:10}) evolves, at late times, to
\begin{equation}
  \label{eq:30}
  \psi_{t} = \frac{|+\rangle \otimes \phi_{+}
  + \epsilon_{t}|-\rangle \otimes \phi_{-}}{\sqrt{1+ \epsilon_t^2}}.
\end{equation}
The evolution of the stochastic quantity $\epsilon_t$ is determined dynamically by the stochastic equation: it either goes to $\epsilon_t \ll 1$, with a probability  $|c_{+}|^2$, or to $\epsilon_t \gg 1$, with a probability 
$|c_{-}|^2$. In the former case, one can say with great accuracy that the state vector has `collapsed' to the definite outcome $|+\rangle \otimes \phi_{+}$ with a probability $|c_{+}|^2$. Similarly, in the latter case one concludes that the state vector has collapsed to $|-\rangle \otimes \phi_{-}$ with a probability $|c_{-}|^2$.
This is how collapse during a quantum measurement is explained dynamically, and random outcomes over repeated measurements are shown to occur in accordance with the Born probability rule. The time-scale over which $\epsilon_t$ reaches its asymptotic value and the collapse occurs can also be computed dynamically. In the present example, for a pointer mass of 1 gram, the collapse time turns out to be about $10^{-4}$ seconds.

Lastly, we can understand how the modified stochastic dynamics causes the outcome of a diffraction experiment in matter wave-interferometry to be different from that in quantum theory. Starting from the fundamental equation (\ref{eq:qmupl1}) it can be shown that the statistical operator $\rho_t = \mathbb{E}[|\psi_t\rangle\langle\psi_t|]$ for a system of $N$ identical particles evolves as
\begin{equation}
\rho_t(x,y) = \rho_0(x,y) e^{- \lambda N (x-y)^2 t/2}.
\end{equation}
Experiments look for a decay in the density matrix by increasing the number of the particles $N$ in an object, by increasing the slit separation $|x-y|$, and by increasing the time of travel $t$ from the grating to the collecting surface.  The detection of an interference pattern sets an upper bound on $\lambda$. The absence of an interference pattern would confirm the theory and determine a specific value for $\lambda$ (provided all sources of noise such as decoherence are ruled out.)

A detailed review of the CSL model and its experimental tests and possible underlying theories can be found in ~\cite{RMP:2012}.
\end{document}